\documentclass{article}

\usepackage{mathptmx}
\usepackage{subfigure}
\usepackage[font=footnotesize]{subfig}
\usepackage[pdftex]{graphicx}  
\usepackage{subfig}
\usepackage{enumitem}
\usepackage[margin=1.3in]{geometry}
\usepackage[parfill]{parskip}
    \usepackage[utf8]{inputenc}

\def\hb{\hbox to 10.7 cm{}}

\date{}

\title{Classifier Selection with Permutation Tests}

\author{
  Marta Arias\\
  \texttt{marias@cs.upc.edu}
  \and
  Argimiro Arratia\\
  \texttt{argimiro@cs.upc.edu}
  \and
  Ariel Duarte-L\'opez\\
  \texttt{aduarte@ac.upc.edu}
}

\begin{document}
  \maketitle

\begin{abstract}
This work presents a content-based recommender system for machine learning classifier algorithms. 
Given a new data set, a recommendation of what classifier is likely to perform best is made based on classifier performance over similar known data sets. This similarity is measured according to a data set characterization that includes several state-of-the-art metrics taking into account physical structure, statistics, and information theory. 
A novelty with respect to prior work is the use of a robust approach based on permutation tests to directly assess whether a given learning algorithm is able to exploit the attributes in a data set to predict class labels, and compare it to the more commonly used $F$-score metric for evaluating classifier performance.
To evaluate our approach, we have conducted an extensive experimentation including 8 of the main machine learning classification methods with varying configurations and 
 65 binary data sets,  leading to over 2331 experiments. Our results show that using the information from the permutation test clearly improves the quality of the recommendations.

\end{abstract}

\section{Introduction}
\label{sec:intro}

The area of machine learning is arguably one of the most popular and successful disciplines in computer science. As a consequence, new algorithms and methods are constantly being designed and developed. Scientists and practitioners who want to apply machine learning solutions to solve a particular problem of interest, could find themselves overwhelmed by the choice of what method to use given the abundance of existing methods. To make matters worse, most algorithms have a number of configurable parameters that need to be adjusted, and unless one understands very well the nature of the parameters it can be a challenge to adjust them well. And even if one is a machine learning expert, searching for optimal models and parameter settings is often highly time-consuming.

The field of \emph{meta-learning} emerged as a way of studying and understanding the performance and behavior of learning algorithms \cite{vilalta2002perspective}. Meta-learning tries to answer important questions such as: Under what circumstances is one learning algorithm better than another? Can we identify scenarios where certain algorithms consistently fail? Meta-learning deals with algorithm recommendation (also called algorithm selection) as well, namely: Is there a way to automatically recommend which algorithm to apply without having to try out hundreds of different variants? Such a recommender system would undoubtedly be of enormous help to the overwhelmed practitioner. 
Our paper focuses on this latter question and presents a classification algorithm recommendation system that, given a previously unseen data set, presents a ranked list of classifiers likely to yield good performance. 

The seminal paper by John Rice \cite{rice1975algorithm}
establishes the basic formalization of the algorithm selection problem.
In Rice's model, the algorithm selection problem comprises three spaces: the problem space, the algorithm space and the performance measure space. The reader will find these three spaces characterized in our present paper: the problem space is represented by our \emph{data set characterization}, the algorithm space by our \emph{classifier characterization}, and finally the performance measure space is represented in two ways: by a permutation test specifically designed to assess how well a classifier exploits its available data \cite{ojala2010permutation}, and by the more common $F$-score metric. Our main contribution is, in fact, the use of this permutation test; as the reader will see, using the permutation test greatly improves the quality of the recommendations made. 
This use of the permutation test stands alone to a large quantity of related work, 
which  focuses on the more common classification measures such as accuracy, precision/recall or $F$-score \cite{vilalta2002perspective,ali2006meta,alexandros2001model,castiello2005meta,tjhimining,moran2009choosing,kopf2000meta,hilario2009data,fernandez2014we}.

To build our recommendation system we need a database to work from. Essentially, our data should contain information on the three spaces described by Rice \cite{rice1975algorithm}: (1) data sets, (2) classifiers including parameter configuration, and (3) performance measures. 
We have designed a database schema that includes tables for (1) the description of many data sets, (2) the description of many classifiers including parameter values used, and (3) performance metrics describing how well a given classifier performed on a particular data set. Such a database has been termed
\emph{experiment database} \cite{vanschoren2012experiment} and is the basis for the evaluation of classifier selection algorithms. In particular, we build a single final \emph{experiment table} whose rows include information on the triples: (data set characterization, classifier and parameter characterization, classification metrics). In this paper, an \emph{experiment} refers to a row of this table.
In Section~\ref{sec:experimental_database} we briefly describe the system that we have developed in order to generate our experiment database and our final experiment table.
This ``final'' experiment table is the starting point of our classifier recommendation engine. 
Given a new data set, our system identifies those experiments in the experiment table corresponding to similar data sets, and builds from these ``similar experiments'' a ranked list of classifiers that worked well for the similar data sets. The working assumption of our recommender system is that the classifiers that worked well in the past for similar data sets should work well for the data set at hand. 
We implement two methods for building our recommendations: one that is based on the $F$-score found in the similar experiments, and a second one which bases the recommendations on the $p$-values obtained using the permutation test from \cite{ojala2010permutation} and described in Section~\ref{subsec:permutation}.
The idea is that the first method ``represents'' the state-of-the-art classifier selection systems all of which use the $F$-score or related standard classification metrics. And the second method represents our proposed idea that permutation tests are valuable tools that are able to capture a different aspect of classifier performance and consequently enriches the process of recommendation.
A detailed account of our recommendation algorithm is given in Section~\ref{subsec:selection}.
Finally, the proposed selection algorithm is evaluated and results obtained are presented in Section~\ref{sec:results}. We close with further theoretical  arguments in favour of permutation test over the F-test.

\paragraph{Related Work.}

The main motivation behind this work is \cite{vanschoren2012experiment}, where authors develop an experimental database.  Besides being conscious of the importance of keeping the information of our research project publicly available,  we had the need to store extra information related with each of the processed experiments, and to control the way of storing the contingency matrix associated to each experiment, in order to analyze and discover the relationship between data sets and classifiers. 
Other set-ups for automatic classifier selection and massive testing of classifier performance can be found in \cite{tjhimining,fernandez2014we}.


As was already mentioned, there exist several approaches for algorithm selection 
\cite{kopf2000meta,castiello2005meta,rendell1990empirical,ali2006meta,alexandros2001model,moran2009choosing,hilario2009data,fernandez2014we,vilalta2002perspective}. These use common content-based recommendation strategies based on different kinds of characterizations of data sets and algorithms.
The closest to our work is \cite{tjhimining},
who study the problem using three different approaches: nearest neighbor search, landmarking by clustering classifiers,
and using a decision tree to organize data sets.
We use the nearest neighbor strategy but in its more general $k$-NN form.
However, the key difference with respect to all previous work is the fact that we do include a test for the goodness of the accuracy metrics, what we believe is a key aspect and, as we show here, helps indeed in improving recommendations.

\section{An experimental database system}
\label{sec:experimental_database}

We have built our own framework   to deal with the massive execution of  experiments and their systematic analysis through the collected statistics.
It is divided in three  steps: 
\begin{enumerate}[leftmargin=*]
\item  Configuration of the experiments: 
 the user  chooses a data set from the database, the classification algorithm, its parameters and the  metric  to evaluate the results. This process can be executed for a single combination of data set and classifier or for multiple combinations.
All the configurations are stored in our database. 
\item  Delivery of the experiment descriptions to the cluster\footnote{All experiments have been carried out in the Computer Science Department's cluster of the Universitat Polit\`ecnica de Catalunya.} where it is processed. In the cluster, the learning algorithm is executed and the value associated to the permutation test is calculated.  
Results are downloaded and stored in our database. 
\item Load the information related to each of the experiments and generate the
experiment table  with the results of the $F$-score and the $p$-value associated to each execution.
\end{enumerate}

For the application development we use Django (version 1.6.1). The database is running over Mysql 5.5.41. 
The cluster used has around 160 physical servers, 1000 CPU cores
and more than 3 TBytes of memory.
For our experiments in classifier selection, we tested eight different classifiers  
over 65 binary data sets. In the following sections we give a brief account of classifiers and data sets used.

\paragraph{Classifiers.}
We have used a popular implementation of eight different classifiers from the data mining framework 
Orange \cite{JMLRdemsar13a}. We have chosen what we believe is a good 
sample of commonly used classification algorithms\footnote{For more detailed information, please consult \cite{tesismaster}. Details omitted from this paper due to lack of space.
}:
 Knn Learner,
Logistic Regression,
 Majority,
 Naive Bayes, Neural Network,
 Random Forest,
 SVM and
 Tree Learner.
Their parameters have been set to realistic and often used values.  

\begin{figure}
  \begin{minipage}[b]{.47\linewidth}
    \centering
    \includegraphics[width=\textwidth]{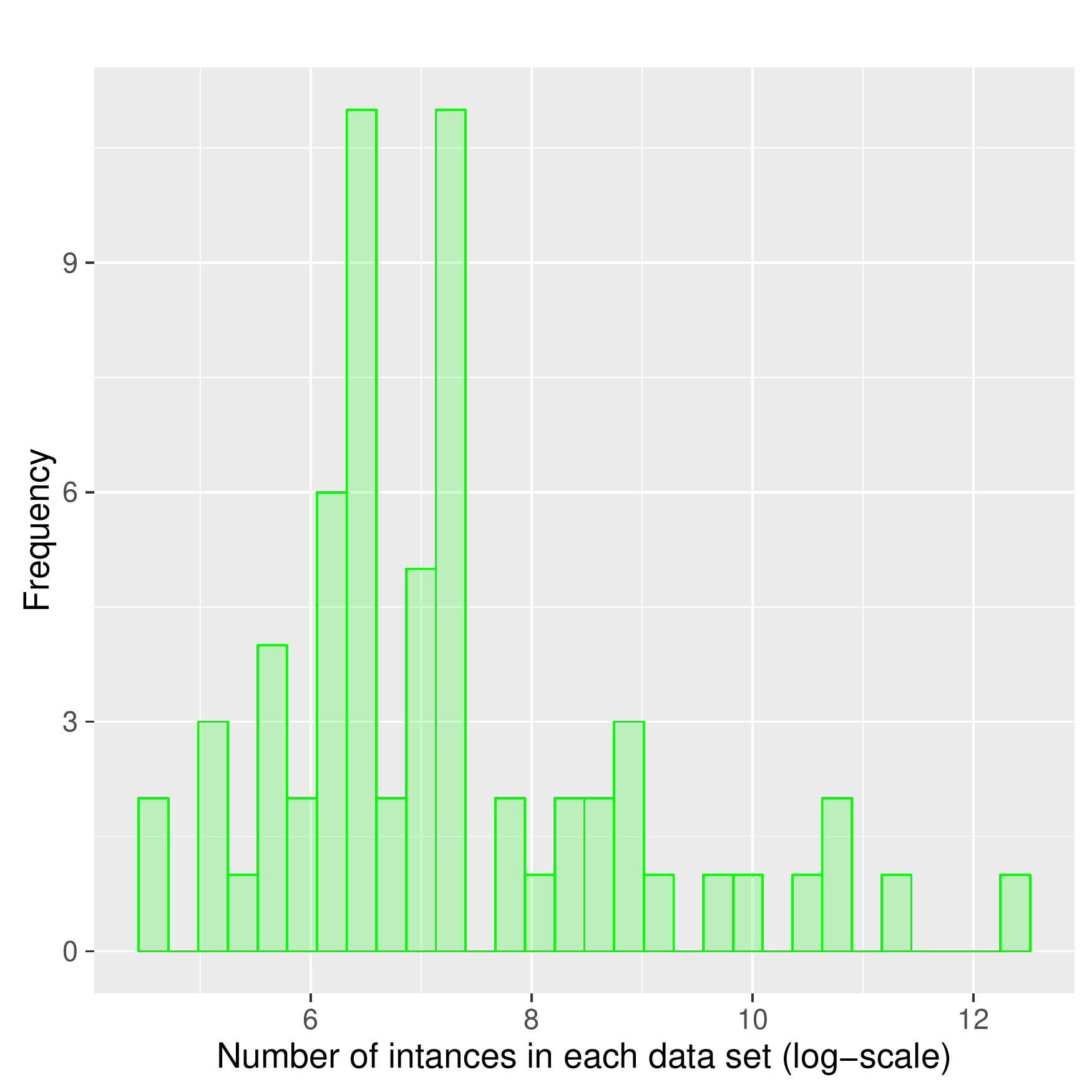}
  \end{minipage}
  \hfill
  \begin{minipage}[b]{.47\linewidth}
    \centering
    \includegraphics[width=\textwidth]{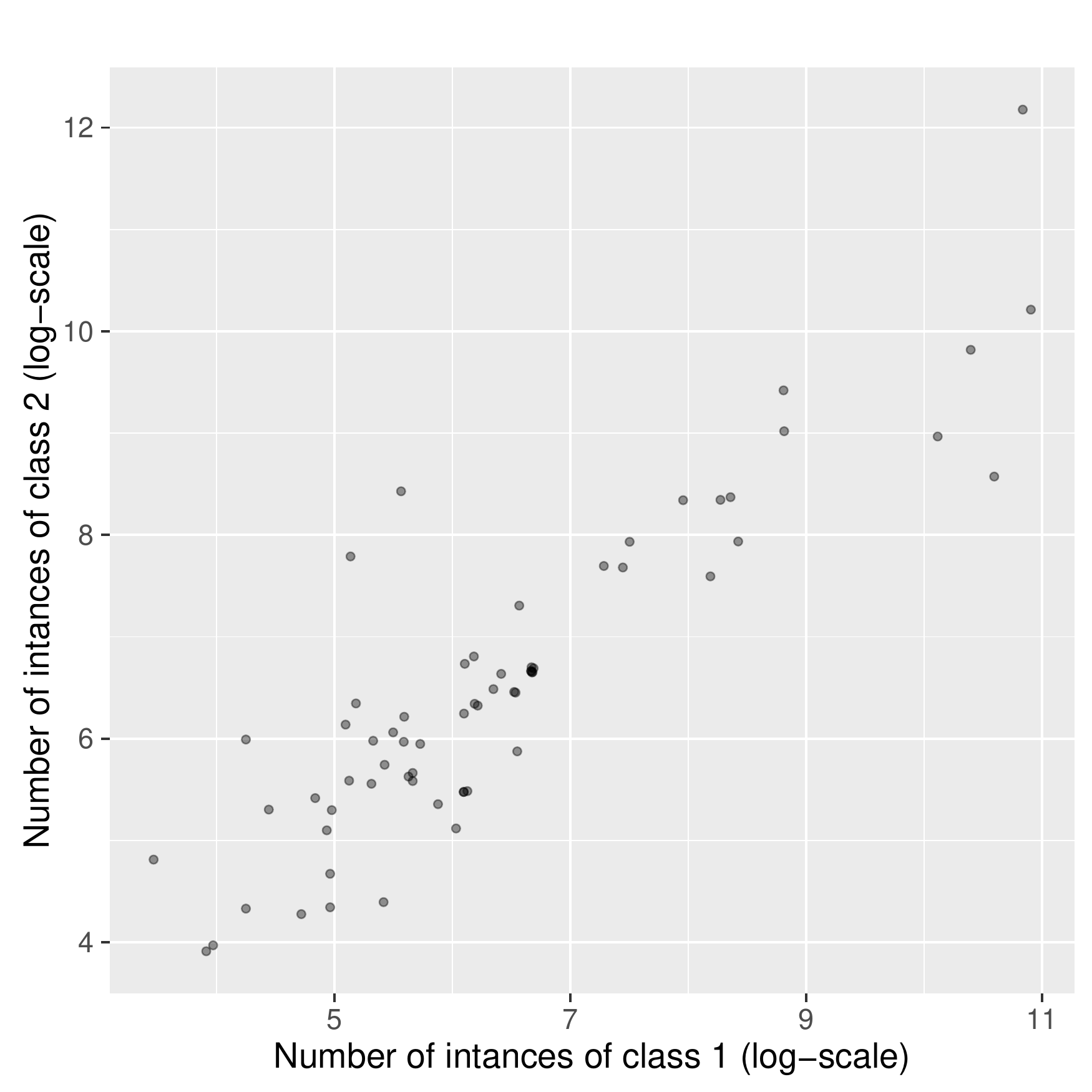}
  \end{minipage}
\caption{Distribution of the size of generated data sets (left) and number of instances per class (right).}\label{fig:splitDS}
\end{figure}

\paragraph{Data sets.}
For simplicity, we focus on binary data sets only. We obtained 35 data sets from the UCI Repository \cite{Lichman:2013}. 
In order to produce some more, we downloaded 15 multi-class data sets and using the one vs. one strategy obtained 30 new binary data sets. In this reduction strategy, we do not use the same
class more than once to avoid dependencies between datasets\footnote{Details on this procedure can be found in \cite{tesismaster}. They have been excluded here for lack of space.}.
Figure~\ref{fig:splitDS} shows that the number of instances per class in each resulting data set is balanced, and covering a wide range of sizes.

\section{Methods}

\subsection{Permutation Tests} 
\label{subsec:permutation}

To assess the performance of classifiers we propose to 
 use a statistical significance test based on permutations. The general  permutation
  testing procedure and its applications are well reviewed  in \cite{good2013}; its particular use  for studying the competence of a classifier is 
examined in \cite{ojala2010permutation}. We follow the latter reference for the definition of this test.

Given an $n\times m$ data matrix $X$, the $i$-th row and $j$-th column of $X$ are denoted by $X_i$ and $X^j$ respectively. Rows represent observations or data points, while columns represent attributes or features. Each data point $X_i$ has a corresponding class label $y_i$, and we let $D= \{(X_i,y_i)\}_{i=1}^n$ be the set of labeled data.  
Let $f$ be the function from data to labels (a classification) learned by a classification algorithm, and let $e(f,D)$ be a test statistic that measures the classifier performance. 
A commonly used  test statistic is
the leave-one-out cross-validation error, defined as
\begin{equation}
e(f,D) = \frac{1}{n}\sum_{i=1}^n I(f_{D\setminus D_i}(X_i) \not= y_i)
\end{equation}
where $f_{D\setminus D_i}$ is the function learned by the classification algorithm by removing the $i$-th observation from the data and $I(\cdot)$ is the indicator function.
 To assess the significance of the values of the error one applies a permutation test, which consist in taking randomized versions of the class labels several times, 
 compute the error statistic for each re-labeling, and consequently obtain a $p$-value  from a null distribution of data samples. 
 
 The null hypothesis of interest is that the data $X$ and the labels $y$ are independent. 
  Then a randomized version $D'$ of $D$ is obtained by permuting the labels. 
 Let  $\hat{D}$ be a set of $k$ randomized versions $D'$ of the original data $D$.  
 Then the empirical $p$-value for classifier $f$ is defined as
\begin{equation}
p = \frac{|D' \in \hat{D}: e(f, D') \leq e(f, D)| + 1}{k + 1},
\label{eq:ePermTest}
\end{equation}

A small $p$-value (we take as threshold 0.05) rejects the null hypothesis that the features and the labels are independent, and consequently the classifier has truly classified and has not found patterns by mere chance (see \cite{ojala2010permutation} for a more in-depth discussion on this permutation test).
In practice, we do Monte Carlo sampling from the set of all permutations to approximate this $p$-value\footnote{Details on how this is done can be found in \cite{tesismaster}.}.

\subsection{Characterization of data sets} 
\label{subsec:dataset}

Our content-based recommendation engine is based on similarity among data sets. In order to assess data set similarity we have compiled a list of 15 data set features. Similarity between data sets is thus defined as the 
Euclidean distance between the corresponding vectors of data set features.
The features that we use for our data set characterization are  taken from \cite{castiello2005meta}, with the exception of   features 3 to 6  which are new in the list below. Their definition is either self-explanatory or included in the list.
The features are categorized according to three aspects: general, statistical and information theoretic. 

It is worth noting that some of these features are global in the sense that they refer to the whole data set, whereas others are computed for each attribute in the data set, and a {\em final feature is given as the average over all attributes}. We mark those features obtained by averaging over all attributes with $[*]$. 

\paragraph{General features:}
Number of instances.
Number of attributes.
Ratio of number of instances to number of attributes.
Has missing values? (binary feature showing whether the data set contains missing values or not).
Missing values (percentage of missing values per attribute) $[*]$.
Unique values (percentage of unique values per attribute) $[*]$.

\paragraph{Statistical features:}
Linear correlation coefficient $[*]$.
Skewness coefficient $[*]$.
Kurtosis coefficient $[*]$.
1-D variance fraction coefficient.

\paragraph{Information-theoretic features:}
Nomalized class entropy.
Normalized attribute entropy $[*]$.
Maximum normalized mutual information.
Equivalent number of attributes.
Noise to signal ratio.

\subsection{Selection procedure} 
\label{subsec:selection}

Given a new data set, the process of classifier selection is outlined as follows:

\begin{enumerate}[leftmargin=*]
\item Load the experiment table

\item Obtain list of 10 classification algorithms by selecting top 2 scoring classifiers from 5 most similar data sets
\item Add the score of the repeated classifiers in the list, return top-scoring classifier
\end{enumerate}
 
To determine similarity among data sets in step 2, the Euclidean distance over the normalized projected vector of data set features is used. The normalization is carried out in the usual way:

\begin{equation}
\tilde{x_i} = \frac{x_i - min(X)}{max(X) - min(X)} 
\label{eq:normalization}
\end{equation}
where $x_i \in X$ and $X$ is the set of values of a data set feature.

For each of the closest neighbors, two best-scoring classifiers are considered. 
When working with $F$-scores, larger values mean better performance.
On the other hand, with $p$-values, smaller values are better. 
After the algorithm decides which are the closest instances, the $score$ associated to each classifier is calculated. 
Equation~\ref{eq:rankF1} shows how the score is calculated when the $F$-scores or the $p$-values are used respectively.
 
\begin{equation}
score = \frac{F1}{(dist_{nei})^2} , \qquad score = \frac{1 - T1}{(dist_{nei})^2}, 
\label{eq:rankF1}
\end{equation}
where $F1$ and $T1$ are the $F$-score and the $p$-value of the best classifier associated to the closest neighbors data set; and $dist_{nei}$ is the distance assigned to the closest data sets.  

There are cases in which a classifier can be selected more than once in the list of the 10 classifiers (two classifiers for each neighbor). We interpret this situation as if the classifier is more prone to give a good performance with the analyzed data set. When this is the case   the $score$ associated to each repeated classifier is added. After processing the $score$ values a ranking is obtained by sorting the algorithms in descending order with respect to their $score$ values. 
We consider the algorithm with the largest $score$ value the one that will produce the best classification on the data set under test.
 
Admittedly, using $k=5$ closest data sets and their 2 best-scoring classifiers is a choice made somewhat arbitrarily.
However, the successful results obtained have encouraged us to continue using these values. 
We recognize that more testing should be done in the future to correctly tune these parameters.

\section{Results} 
\label{sec:results}

Our results are divided into two main sections.
Section~\ref{subsec:F1vsT1} compares the $F$-scores
and the $p$-values obtained from the permutation test. 
The reason to include this statistical
information about the two metrics is for the reader to understand what values these metrics take across our
experiment table. 
As a quick reminder, each row of our experiment table contains information on a particular data set, on a particular classifier, and includes the values obtained by these two metrics 
($F$-score and $p$-value) 
when applying the classifier to the data set represented in that row.

The second part of this section shows the results obtained by our two selection algorithms, namely, the one that bases its recommendation on $F$-scores and the one that bases its recommendation on $p$-values. Section~\ref{subsec:evaluation} first explains the metric that we have used for the system's evaluation and results obtained then follow.

\subsection{$F$-scores vs. $p$-values}
\label{subsec:F1vsT1}

\begin{figure}[!t]
  \begin{minipage}[b]{.47\linewidth}
    \centering
    \includegraphics[width=\textwidth]{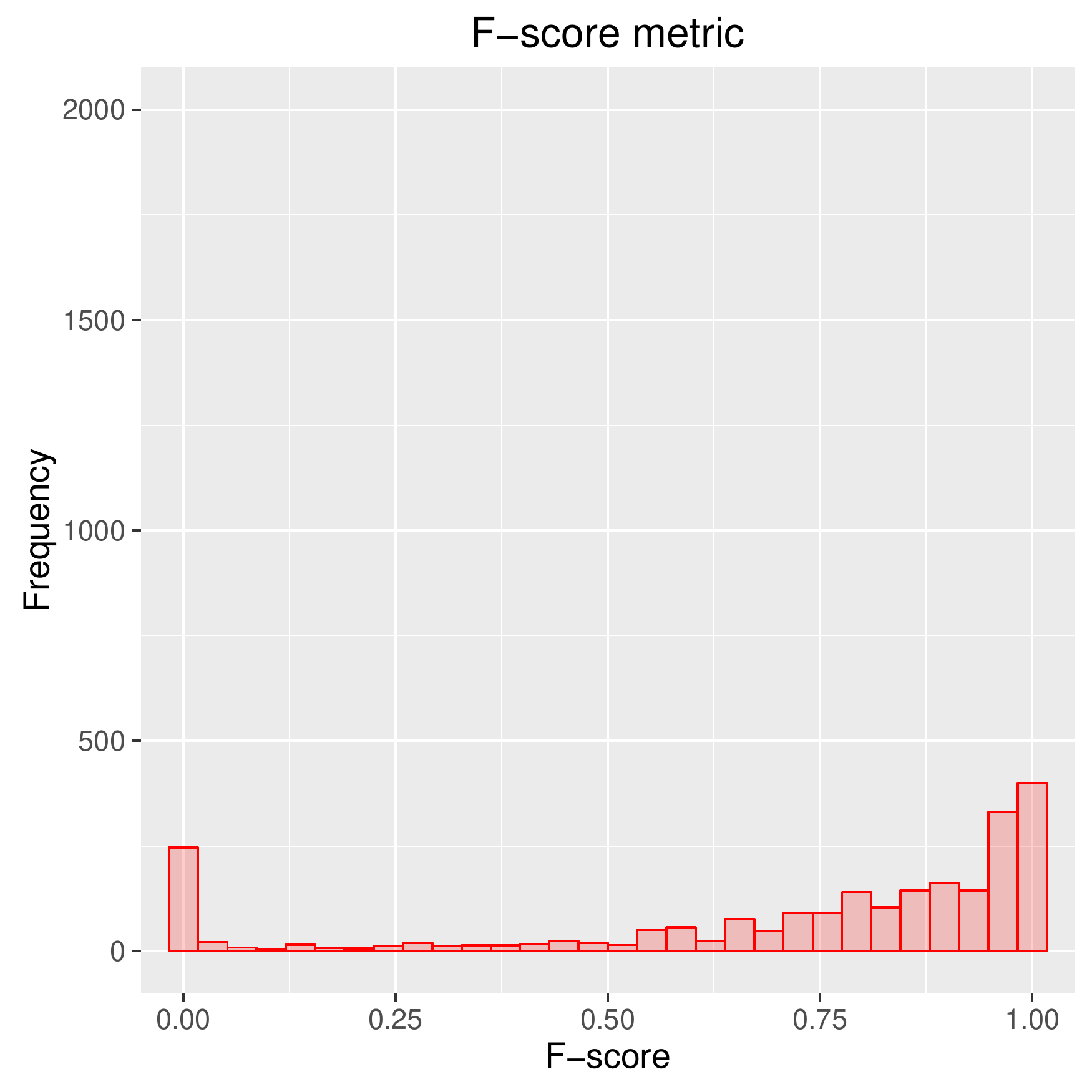}
  \end{minipage}
  \hfill
  \begin{minipage}[b]{.47\linewidth}
    \centering
    \includegraphics[width=\textwidth]{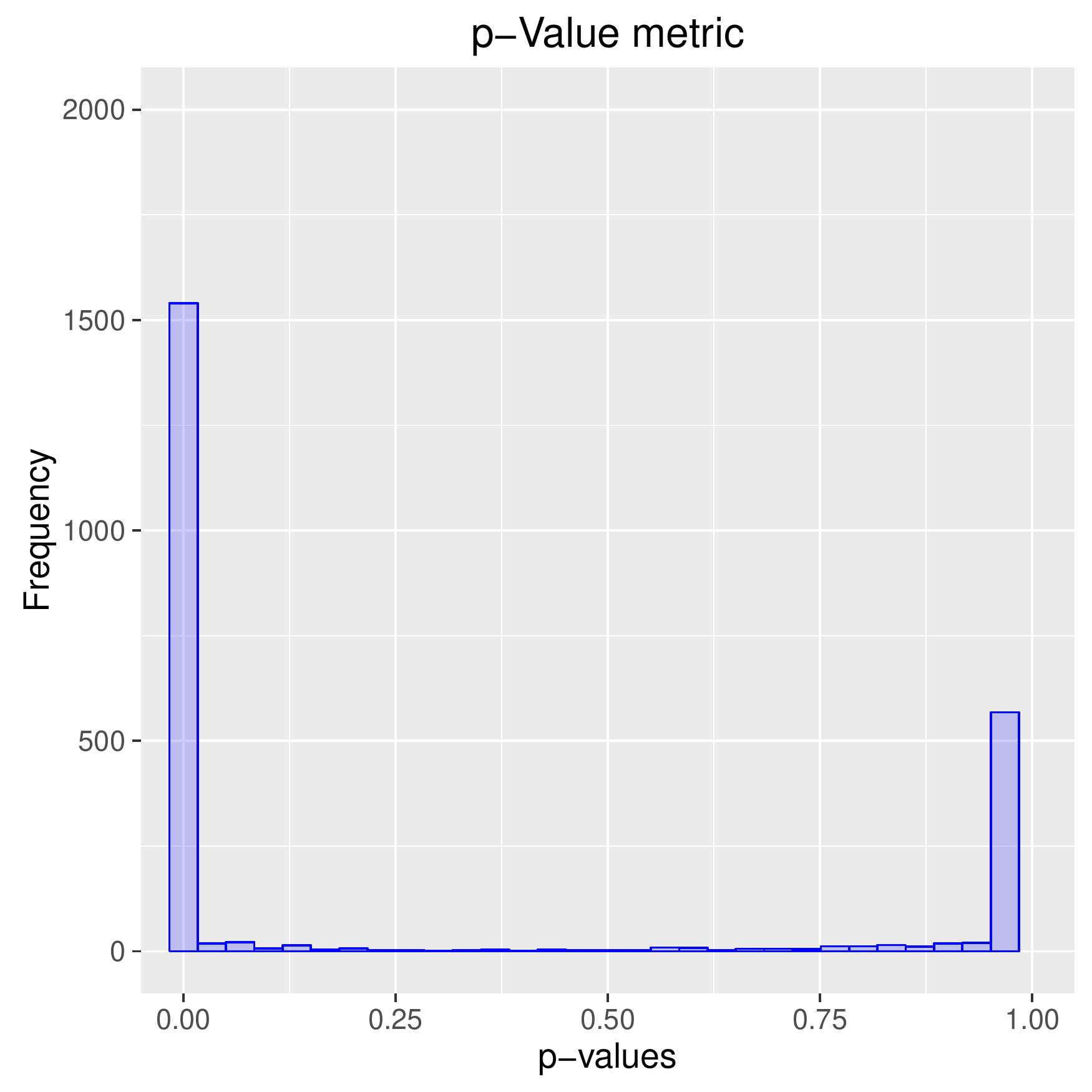}
  \end{minipage}
 \caption{Distribution of the $F$-scores and the $p$-values.}
 \label{fig:distrib}
\end{figure}

Figure~\ref{fig:distrib} shows the distribution of the values obtained by these two metrics in our experiment table. Note that high $F$-score values and low $p$-values mean better classification performance.  
It is apparent that the distributions are very different and so it is safe to conclude that
these two metrics are capturing different aspects of the classification process. In particular, $p$-values seem to divide the experiments into two well-differentiated groups: those classifiers that are able to exploit the data present in the attributes of the data set (left bump in the distribution) and those that are not (right bump). The former group contains the clear majority of the experiments, thus concluding that, according to our permutation test of Section~\ref{subsec:permutation}, learning does takes place in the majority of the experiments.
$F$-scores are in contrast more uniformly distributed in the sense that there is a significant fraction of experiments taking values in the mid section. Both distributions agree, however, that the majority of experiments conducted (present in our experiment table) are able to learn since these distributions are both biased to the ``good side'' of the spectrum of values.

Next, we study directly the correlation between values of ($F$-score, $p$-values) observed in our experiment table. 
To better understand this relation, we have discretized the values for each metric into three groups: good behavior, neutral, and poor behavior. The threshold values that we use for the discretization are shown in Table \ref{tab:intervals}

This discretization partitions the relation $F$-score vs. $p$-value into nine regions. 
Table \ref{tab:discMet} shows the number and percentage of experiment rows that fall into each of these nine discretized regions. As can be seen in the table the three regions where both metrics agree (good/good, neutral/neutral, and poor/poor) account for the majority of the experiments (50.3\%), but there is a significant proportion of experiments
where the two metrics disagree (remaining  49.7\%). 
Moreover, the $F$-score statistic seems to be more demanding since in  7.8\%  of the experiments $F$-score rates poorly whereas $p$-value rates well. The opposite ($p$-value rates poorly whereas $F$-score rates well) only happens in  0.6\%  of the experiments. Note, however, that these percentages are small so this distinction may not be significant.
In addition, these percentages depend on the thresholds, which, admittedly, are somewhat arbitrary albeit reasonable.

\begin{table}
  \begin{minipage}[b]{.47\linewidth}
\centering
\caption{Threshold values for each metric}
\begin{tabular}{cccc}
\hline\noalign{\smallskip}
 & Good & Neutral & Poor \\
\noalign{\smallskip}
\hline
\noalign{\smallskip}
F-scr & $[0.9; 1]$ & $[0.5; 0.9)$ & $[0; 0.5)$ \\
$p$-val & $[0;0.045]$ & $(0.045; 0.2]$ & $(0.2; 1]$ \\
\hline
\end{tabular}
\label{tab:intervals}
 \end{minipage}
  \hfill
  \begin{minipage}[b]{.47\linewidth}
\centering
\caption{Number and percentage of experiment rows falling into each of the nine discretized regions.}
\begin{tabular}{cccc}
\hline\noalign{\smallskip}
\textsl{$F$-score} & \textsl{$p$-value} & \textsl{Count}  & \textsl{ \%}\\
\noalign{\smallskip}
\hline
\noalign{\smallskip}
poor & poor & 234& 10\\
poor & neutral & 32&1.4\\
poor & good & 182 & 7.8\\
neutral & poor & 471&20.2\\
neutral & neutral & 20&0.9\\
neutral & good & 459&19.7\\
good & poor & 13&0.6\\
good & neutral & 2&0.1\\
good & good & 918&39.4\\
%
\hline
\end{tabular}
\label{tab:discMet}
 \end{minipage}\end{table}

\subsection{Evaluation of classifier selection algorithm} 
\label{subsec:evaluation}

Here, we evaluate the results obtained by our two approaches for recommending classifiers (using $F$-scores and using $p$-values).
We propose the following evaluation method, which is known as \emph{leave one out cross validation} in machine learning. That is, 
for each data set in our experiment table, we rank classifiers according to our algorithm of Section~\ref{subsec:selection} using the portion
of the experiment table  \emph{involving other data sets only}. For a given data set $i$, we collect the recommendation made by the
selection algorithm; let this recommended classifier be $c_i$. Then, we look into the rows of the experiment table that involve data set $i$ and locate the
recommended classifier $c_i$. We compute the position that $c_i$ occupies according to the $F$-score metric present in the experiment rows. Let $rank_i$
be the position occupied by $c_i$ after sorting classifiers according to $F$-score in descending order. Note that  rank positions with low values indicate that the
recommended classifier performed well (relative to others), whereas rank positions with high values indicate that the recommended classifier performed poorly.

\begin{figure}[!t]
  \begin{minipage}[b]{.47\linewidth}
    \centering
    \includegraphics[width=\textwidth]{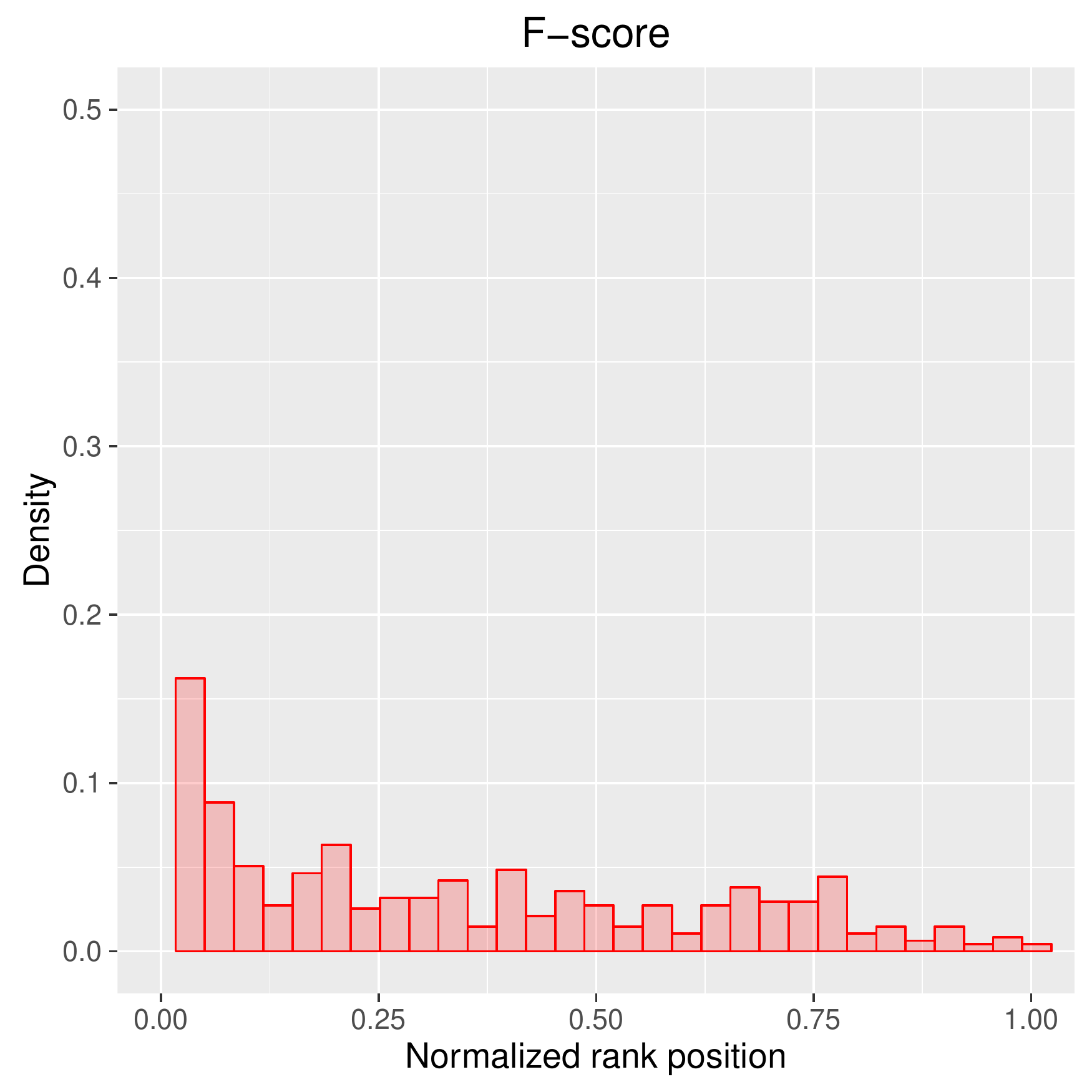}
  \end{minipage}
  \hfill
  \begin{minipage}[b]{.47\linewidth}
    \centering
    \includegraphics[width=\textwidth]{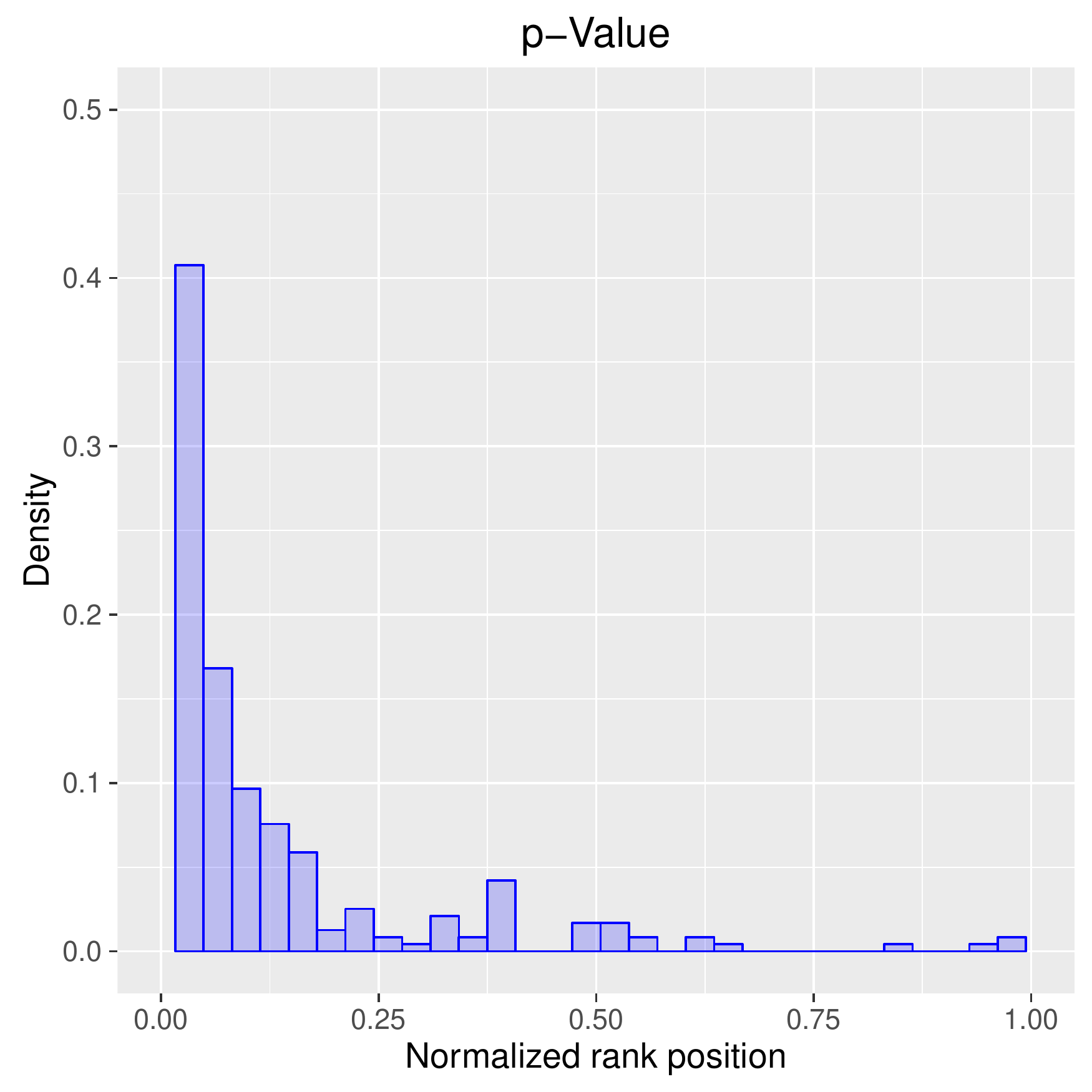}
  \end{minipage}
 \caption{Distribution of the normalized rank positions for the recommendation strategy using $F$-score, and for the recommendation strategy using $p$-value.}
 \label{fig:exp_histograms}
\end{figure}

\begin{figure}[!t]
    \centering
    \includegraphics[width=0.47\textwidth]{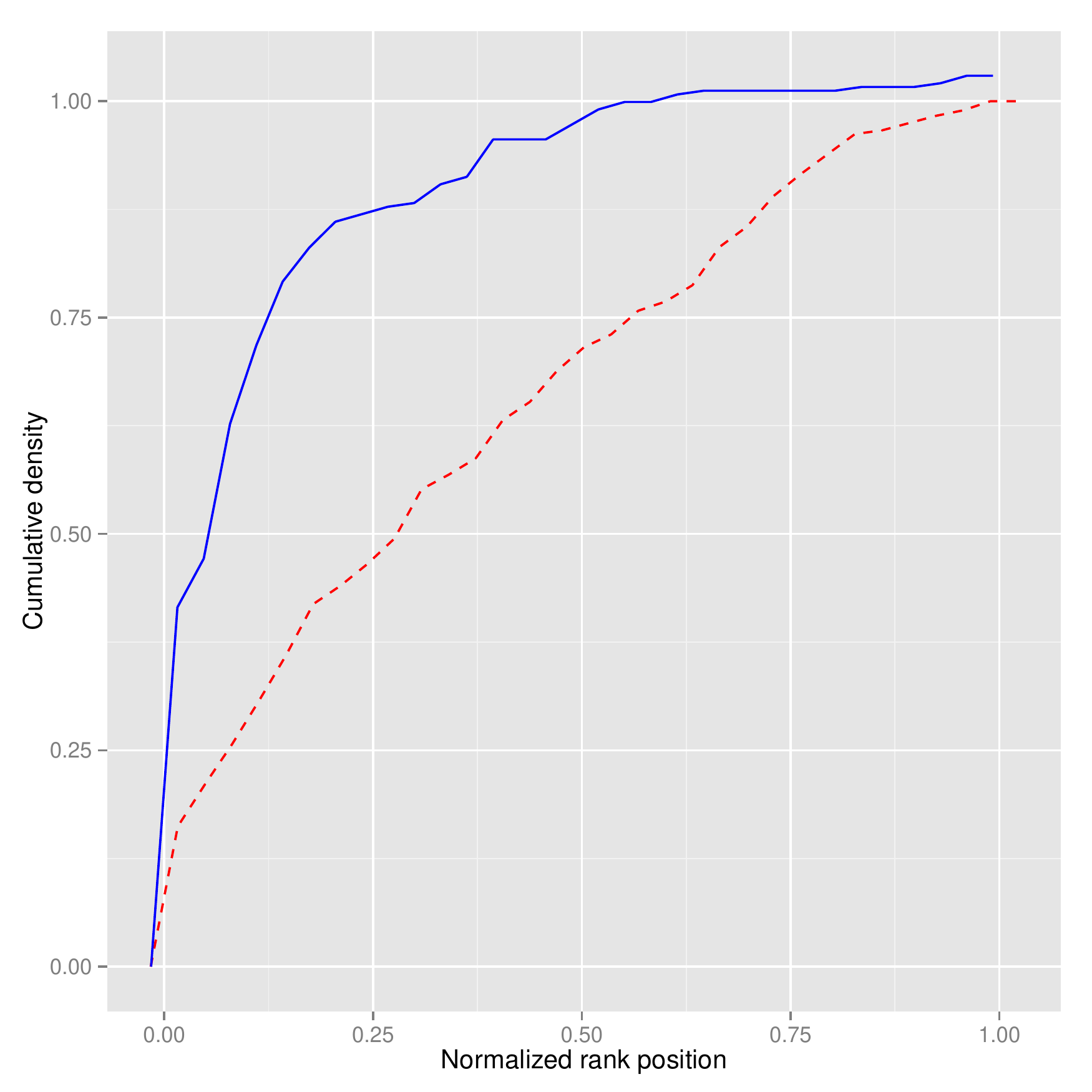}
    \caption{Cumulative density plots for $F$-score (dashed red line) and $p$-values (continuous blue line).}
    \label{fig:rocs}
\end{figure}

Since not all classifiers are applicable to all data sets\footnote{This depends on the nature of attributes present in the data set and the classifier, e.g., not all classifiers can work on categorical data.}, the number of competing classifiers can differ for each data set, and thus normalizing by classifier list length is needed in order to make the rank positions comparable and compute an aggregate over all $rank_i$ for each data set $i$. Thus, let $nrank_i$ be the normalized rank position of recommended classifier for data set $i$, which is computed as $rank_i / m_i$, where $m_i$ is the number of experiments using data set $i$. This value ranges from 0 to 1, with 0 being perfect recommendation and 1 being worst recommendation possible.
Figure~\ref{fig:exp_histograms} shows 
the improved quality of the recommendations when using $p$-values.
To summarize the results for both strategies, 
we perform area under the curve (AUC) analysis on the cumulative density plots for both histograms. Figure~\ref{fig:rocs} shows the cumulative density for both strategies. The AUC for the strategy using $p$-values is  0.9353, indeed far superior to 0.6699   obtained when using $F$-scores. The main result of this paper is thus confirmed.

\section{Conclusions} 
In this paper we have presented a novel strategy for the algorithm selection problem in the context of binary classification.
Our approach is
based on a permutation test that directly captures whether a classifier exploits data in its predictions. Our results
confirm the superiority of using such a permutation test over the more common approach of using classification performance metrics such as $F$-scores.

As a closing remark we offer a brief explanation for regarding a permutation test as a more precise classifier performance evaluator than the $F$-score.
The $F$-score is known to be a biased measure in that it ignores the performance in correctly handling negative examples and propagates  the  underlying  marginal  prevalences  and  biases \cite{powers}.
A permutation test considers all classified examples and gives an  approximation to the exact distribution of the errors in classification: if we were able to do all permutations we would get the exact distribution, regardless of parameters \cite{good2013}. 
Hence, one can always sharpen the ranking by performing as many permutations as time allows.

\section*{Acknowledgments}  
{ M. Arias and A. Arratia acknowledge support of MINECO project APCOM (TIN2014-57226-P),  
and Gen. Cat. SGR2014-890 (MACDA). We thank the \emph{RDLab} of the Computer Science Dept. at UPC for their support in using the cluster.}

\end{document}